\documentclass[reprint,aps]{revtex4-1}
\usepackage{amsmath}
\usepackage{amsfonts}
\usepackage{graphicx}
\usepackage{mathdots}
\usepackage{xcolor}
\usepackage{xspace}
\usepackage[toc,page]{appendix}

\begin{document}

\title{Wave-packet revival in a Floquet engineering quadratic potential
system}
\author{J. Cao}
\email{These authors contributed equally: J. Cao, H. Shen}
\author{H. Shen}
\email{These authors contributed equally: J. Cao, H. Shen}
\author{R. Wang}
\email{wangr@tjnu.edu.cn}
\author{X. Z. Zhang}
\email{zhangxz@tjnu.edu.cn}
\address{College of Physics and Materials Science, Tianjin Normal
University,Tianjin 300387, China}
\begin{abstract}
We investigate the quantum dynamics of a one-dimensional tight-binding
lattice driven by a spatially quadratic and time-periodic potential. Both
Hermitian ($J_1 = J_2$) and non-Hermitian ($J_1 \neq J_2$) hopping regimes
are analyzed. Within the framework of Floquet theory, the time-dependent
Hamiltonian is mapped onto an effective static Floquet Hamiltonian, enabling
a detailed study of the quasi-energy spectrum as function of the driving
frequency $\omega$. By applying a gauge transformation, we find that
critical frequencies $\omega_c$ emerge, at which nearly equidistant
quasi-energy ladders appear, as revealed by a pronounced minimum in the
normalized variance $\Delta(\omega)$ of the level spacings. This spectral
regularity leads to robust periodic revivals and Bloch-like oscillations in
the time evolution. Numerical simulations confirm that such coherent
oscillations persist even in the non-Hermitian regime, where the periodic
driving stabilizes an almost real and uniformly spaced quasi-energy ladder.
\end{abstract}

\maketitle

\section{Introduction}

Periodically driven (Floquet) quantum systems provide a versatile toolbox
for engineering effective Hamiltonians and dynamical properties inaccessible
in static settings. By modulating parameters in time, one can synthesize
novel band structures, induce topological phases, renormalize interaction
strengths, and produce engineered steady states; these techniques have been
applied across diverse platforms, including cold atoms, photonics, and
condensed matter \cite%
{Bara,Muell,Yao,Liu,Burger,Pan,Recht,Szame,QQC,Zhen,Shu,WHF,Oka,Sondhi,Liang,Mori,Upen,LJF,WH,YXM,MG}%
. Practically, Floquet engineering is attractive because it trades hardware
complexity for temporal control: Rather than fabricating a different static
lattice for each target Hamiltonian, one can program desired effective
dynamics via driving protocols. Conceptually, Floquet systems blur the
distinction between static and dynamical symmetries and open avenues to
explore genuinely time-dependent phases of matter with no equilibrium
analogue.

Despite this promise, time-periodic drives also introduce unique challenges.
A central issue is heating: Continuous energy absorption from the drive can
erase engineered features and, in isolated many-body systems, eventually
lead to a featureless infinite-temperature state. A common mitigation
strategy is to operate in regimes (e.g., high-frequency expansions,
prethermal windows, or many-body localized backgrounds) where energy
absorption is parametrically slow, allowing the engineered Floquet
Hamiltonian to govern long-lived transient dynamics. Within this controlled
Floquet regime, however, a host of nontrivial and useful dynamical phenomena
can be realized, including coherent revivals, Floquet-induced localization,
and photon-assisted band reconstruction. These effects make Floquet
platforms particularly suitable for quantum-state control and for realizing
time-dependent spectroscopic probes.

In parallel, non-Hermitian (\textrm{NH}) quantum systems, effectively
described by Hamiltonians that break Hermiticity through, e.g., asymmetric
hopping, gain/loss terms, or postselected dynamics, have emerged as a rich
platform for novel spectral and dynamical phenomena \cite%
{Bender1,Bender2,Ganainy}. Non-Hermitian models naturally describe open,
driven-dissipative platforms such as photonic lattices, active
metamaterials, and certain electronic circuits. They support unique features
including complex-energy band structures, exceptional points \cite%
{Dembo,WD,WR,Miri,Berg}, the non-Hermitian skin effect \cite%
{GJB,Longhi1,Sato,FC,ChenYF,LQH}, and sensitivity-enhanced responses \cite%
{Wier,YangL,Hodaei,ZhaiH,ChenS}. Integrating non-Hermitian physics with
Floquet driving thus raises critical questions: Can periodic driving sculpt
non-Hermitian spectra, stabilize desired real subsectors, or enable
dynamical control of gain/loss? Conversely, can non-Hermitian mechanisms be
harnessed to expand the repertoire of Floquet engineering?

A potential objection is that Floquet drives, which already tend to heat
isolated systems, would only have their engineered features destroyed more
rapidly by the gain or loss mechanisms inherent to non-Hermitian physics.
This intuition, however, is incomplete. First, many experimental
realizations of Floquet engineering (e.g., photonic waveguides, driven
cavity arrays, cold-atom experiments with controlled losses) are
intrinsically open or driven-dissipative, where non-Hermiticity must be
treated on an equal footing with the periodic drive. Second, non-Hermiticity
is not purely detrimental; when combined with temporal modulation, it can
lead to the dynamical stabilization of spectral subsectors, effectively
suppressing net gain/loss within a targeted manifold. In other words, the
interplay between driving and non-Hermiticity can create and protect useful
dynamical behavior, such as long-lived quasi-Hermitian ladders, rather than
simply accelerating decoherence.

Motivated by these possibilities, in this work we study a paradigmatic
one-dimensional tight-binding chain subjected to a spatially quadratic and
time-periodic on-site potential. Our model interpolates between Hermitian
and non-Hermitian regimes via asymmetric nearest-neighbor hopping amplitudes
($J_{1}$ and $J_{2}$), capturing experimentally relevant situations such as
cold-atom lattices with modulated potentials \cite%
{ZSL,Garbe,ZhouS,Lobser,Price,Ali,ZHP} or photonic arrays with designed
curvature and nonreciprocity \cite{MAB,ChenY,CWZ,Longhi2,WS}. Within the
framework of Floquet theory, we map the time-dependent problem onto an
effective static Floquet Hamiltonian in Sambe space and analyze how the
quasi-energy spectra of Floquet eigenstates evolve with the driving
frequency $\omega $.

Our main findings are as follows. At high drive frequencies, the system is
well described by the time-averaged Hamiltonian and supports extended
Bloch-like states. In the limit of vanishing frequency ($\omega \rightarrow
0 $), the static quadratic confinement dominates, and eigenstates are
strongly localized, forming harmonic-oscillator-like and Wannier-Stark-like
families. The condition for the emergence of the Floquet ladder structure
(nearest-neighbor quasi-energy spacings) is analytically obtained through a
gauge transformation, and is numerically identified by a pronounced minimum
in the normalized variance $\Delta (\omega )$ of nearest-neighbor
quasi-energy spacings. Dynamically, this equidistant structure gives rise to
robust Bloch-like revivals: An initial wave packet supported on the ladder
subspace exhibits periodic return at times $t_{c}=2\pi /\Delta E$, where $%
\Delta E$ is the ladder spacing. Importantly, we find that in the
non-Hermitian regime, the periodic drive can dynamically stabilize an almost
real, uniformly spaced quasi-energy ladder, i.e., a near-Hermitian
subsector, thereby enabling Hermitian-like revivals even when the underlying
static Hamiltonian lacks Hermiticity. Our analysis combines analytical
Floquet arguments (Sambe space block structure, high-frequency averaging,
and perturbative hybridization) with systematic numerical diagonalization of
truncated Floquet matrices, spectral diagnostics (spacing variance), and
time-dependent wave-packet propagation. These complementary tools allow us
to identify the frequency windows where ladder formation is robust and to
verify the resulting revival dynamics.

The remainder of the paper is organised as follows. In Sec. \ref{Model
Hamiltonian}, we introduce the model Hamiltonian for a spatially quadratic
and time-periodic potential and discuss its behavior in both the static and
high-frequency limits. Sec. \ref{Floquet Hamiltonian and Solutions} is
devoted to the Floquet Hamiltonian and its solutions under finite driving
frequencies. In Sec. \ref{Gauge Transformation}, we present the frequency
condition for the nearly equidistant quasi-energy structure in the Floquet
spectrum, derived via a gauge transformation. Sec. \ref{Bloch oscillations}
demonstrates that a finite driving frequency induces nearly equidistant
quasi-energies in both Hermitian and non-Hermitian regimes. Finally, we
summarize the key results and outline future research directions in Sec. \ref%
{Summary}.

\section{Model Hamiltonian}

\label{Model Hamiltonian}

We consider a one-dimensional tight-binding lattice driven by a spatially
quadratic, time-periodic potential, described by the Hamiltonian
\begin{equation}
\mathcal{H}(t)=H_{0}+H(t),  \label{H_og}
\end{equation}%
where the static hopping term reads
\begin{equation}
H_{0}=-\sum_{l=-L}^{L}\left( J_{1}a_{l+1}^{\dagger
}a_{l}+J_{2}a_{l}^{\dagger }a_{l+1}\right) ,  \label{H0_nonH}
\end{equation}%
and the time-dependent on-site potential is given by
\begin{equation}
H(t)=\sum_{l=-L}^{L}F(l,t)n_{l},\quad n_{l}=a_{l}^{\dagger }a_{l}.
\label{Ht}
\end{equation}%
Here $a_{l}^{\dagger }$ ($a_{l}$) creates (annihilates) a spinless fermion
at site $l$. The parameters $J_{1}$ and $J_{2}$ denote the right- and
left-hopping amplitudes, respectively. The parameter $L$ determines the
lattice size. The model is Hermitian and reduces to the standard
nearest-neighbor hopping Hamiltonian when $J_{1}=J_{2}=J\in \mathbb{R}$. In
contrast, for $J_{1}\neq J_{2}$, Hermiticity is broken and the system
describes a nonreciprocal (non-Hermitian) tight-binding chain, which can be
realized in photonic lattices with asymmetric couplings or synthetic
nonreciprocal waveguide systems.

\begin{figure}[tbp]
\centering
\includegraphics[bb=61 10 753 273, width=8.5cm, clip]{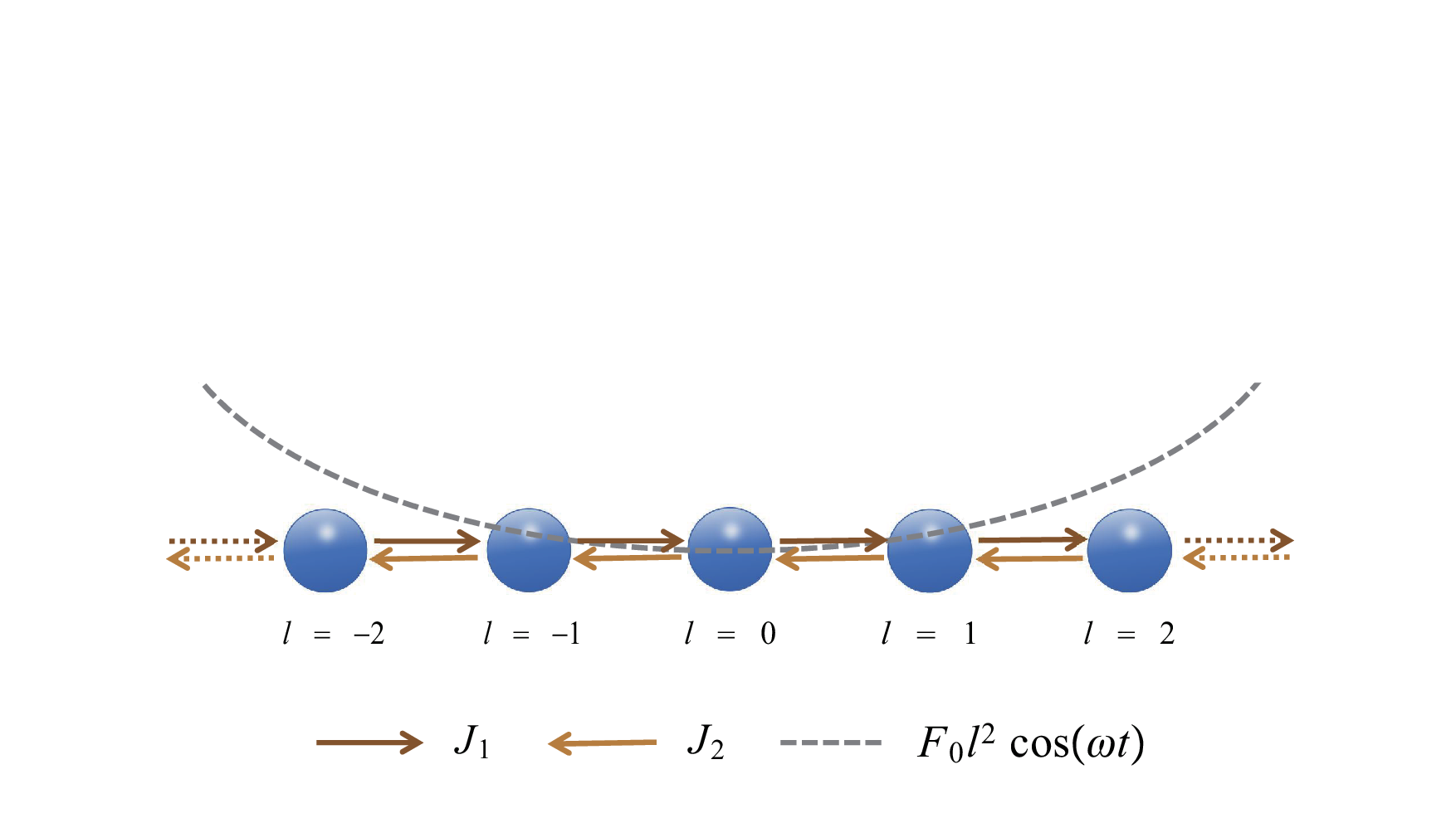}
\caption{Schematic illustration of the quadratic Floquet Hamiltonian $%
\mathcal{H}(t)$. The blue spheres represent lattice sites, the brown arrows
indicate asymmetric hoppings $J_{1}$and $J_{2}$, the gray dashed curve shows
the time-dependent on-site potential $F_{0}l^{2}\cos (\protect\omega t)$.
Such a temporally modulated harmonic confinement can be realized in
ultracold-atom systems by periodically varying the trapping frequency
through the laser intensity or magnetic-field gradients, or in photonic
waveguide arrays by engineering a periodically modulated curvature along the
propagation axis.}
\label{fig1}
\end{figure}

The driving field is spatially quadratic and temporally harmonic,
\begin{equation}
F(l,t)=F_{0}l^{2}\cos (\omega t),
\end{equation}%
where $F_{0}$ controls the curvature of the potential and $\omega $ denotes
the driving frequency. The schematic structure of $\mathcal{H}(t)$ is
illustrated in Fig. \ref{fig1}. The blue spheres represent lattice sites,
the brown arrows indicate asymmetric hoppings $J_{1}$\ and $J_{2}$, and the
gray dashed curve shows the time-dependent quadratic modulation centered at $%
l=0$. Such a temporally modulated harmonic confinement can be realized in
ultracold-atom systems by periodically varying the trapping frequency
through the laser intensity or magnetic-field gradients, or in photonic
waveguide arrays by engineering a periodically modulated curvature along the
propagation axis.

\subsection*{Static limit ($\protect\omega = 0$)}

In the absence of driving, Eq.~(\ref{H_og}) reduces to
\begin{equation}
H_{1}=-\sum_{l=-L}^{L}(J_{1}a_{l+1}^{\dagger }a_{l}+J_{2}a_{l}^{\dagger
}a_{l+1})+\sum_{l=-L}^{L}F_{0}l^{2}a_{l}^{\dagger }a_{l}.  \label{H1_nonH}
\end{equation}%
This represents a quadratic confinement superimposed on a uniform hopping
lattice. For the Hermitian case ($J_{1}=J_{2}=J$), the quadratic potential
breaks the translational invariance and induces an asymmetric, nonlinear
energy-level structure. The low-energy eigenstates are confined near the
potential minimum around $l=0$, resembling harmonic-oscillator-like modes,
while the high-energy eigenstates localize near the boundaries, forming
parity-related pairs of Wannier-Stark-like localized states \cite{Ali}.
These two types of eigenstates originate from the competition between the
hopping amplitude $J$ and the confinement strength $F_{0}$: When $J\gg F_{0}$%
, the eigenstates remain extended and approximately harmonic, whereas for
large $F_{0}$, the eigenstates become strongly localized away from the
center.

\begin{figure}[tbp]
\centering
\includegraphics[bb=64 63 1262 1047, width=9 cm, clip]{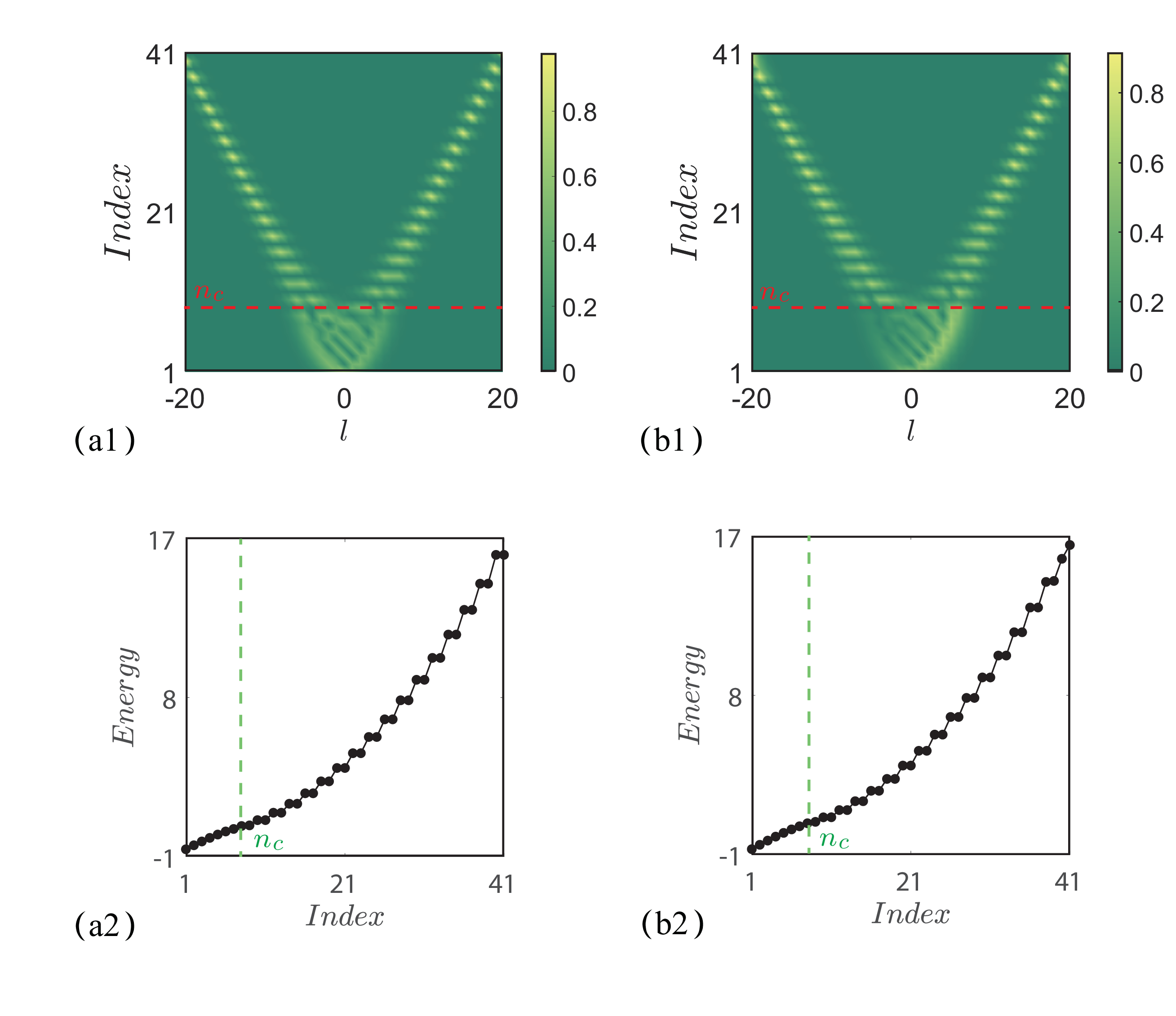}
\caption{Numerical results for the eigenenergies and eigenstates of the
quadratic systems ($\protect\omega =0$) as a function of the site index and
energy index with Hermitian case ($J_{1}=J_{2}$) and non-Hermitian case ($%
J_{1}\neq J_{2}$). The green dashed lines in panels (a2-b2) mark the
crossover index ($n_{c}$) separating harmonic-oscillator-like states [below
the red dashed lines in panels (a1-b1)] from Wannier-Stark-like states
(above it). The states lying below the red dashed line in panel (b1) exhibit
the non-Hermitian skin effect. The parameters are (a1-b1) $J_{2}=0.35$, and
(a2-b2) $J_{2}=0.5$. Other system parameters are $L=20$, $J_{1}=0.35$, and $%
F_{0}=0.04$. The colorbar indicates the probabilities at each position in
the eigenstates.}
\label{fig2}
\end{figure}

For the non-Hermitian case ($J_{1}\neq J_{2}$), the asymmetric hopping
introduces directional localization and complex eigenenergies, producing a
non-Hermitian skin effect in which the eigenstates accumulate toward one
boundary. The quadratic potential counteracts this nonreciprocal
localization by energetically favoring the lattice center. The resulting
eigenstates exhibit a crossover from skin-localized edge modes to confined
bulk-like states as $F_{0}$ increases. Figure \ref{fig2} illustrates typical
eigenstates and eigenenergies of $\mathcal{H}(0)$\ in both Hermitian case ($%
J_{1}=J_{2}$) and non-Hermitian case ($J_{1}\neq J_{2}$) quadratic systems ($%
\omega =0$). The green dashed lines in Figs. \ref{fig2}(a2-b2) mark the
crossover index ($n_{c}$) separating harmonic-oscillator-like states [below
the red dashed lines in Figs. \ref{fig2}(a1-b1)] from Wannier-Stark-like
states (above it). The states lying below the red dashed line in Fig. \ref%
{fig2}(b1) exhibit the non-Hermitian skin effect. Increasing $F_{0}$\
reduces the number of extended central modes and enhances localization near
the edges. Specifically: (a1-b1) $J_{2}=0.35$; (a2-b2) $J_{2}=0.5$. Other
system parameters are $L=20$, $J_{1}=0.35$, and $F_{0}=0.04$. The colorbar
indicates the probabilities at each position in the eigenstates.

\subsection*{High-frequency limit ($\protect\omega \to \infty$)}

In the opposite limit of infinitely fast driving, the time-dependent
potential oscillates so rapidly that its time average over one period
vanishes. Expanding the Floquet Hamiltonian to leading-order in high
frequency, we obtain
\begin{equation}
H_{\mathrm{eff}}=\frac{1}{T}\int_{0}^{T}\mathcal{H}(t)dt=H_{0},\quad T=\frac{%
2\pi }{\omega }.
\end{equation}%
Hence, the system effectively behaves as a uniform tight-binding chain with
hopping amplitudes $J_{1}$ and $J_{2}$. In the Hermitian limit, this
restores full translational symmetry and leads to extended Bloch eigenstates
with dispersion $E(k)=-2J\cos k$. In the non-Hermitian case, the effective
Hamiltonian retains asymmetric hopping and exhibits a complex energy
spectrum of the form
\begin{equation}
E(k)=-J_{1}e^{ik}-J_{2}e^{-ik},
\end{equation}%
forming an elliptical loop in the complex plane. The associated eigenstates
remain extended under periodic boundary conditions but display boundary
accumulation (skin localization) under open boundary conditions.

\subsection*{Physical crossover between the two limits}

The two limits, $\omega \rightarrow 0$ and $\omega \rightarrow \infty $,
define opposite dynamical regimes of the driven system. In the adiabatic
limit ($\omega \rightarrow 0$), the system follows the instantaneous
eigenstates of $\mathcal{H}(0)$, exhibiting quasi-static confinement
dominated by the quadratic potential and, in the non-Hermitian case, a
competition between confining localization and asymmetric hopping. In the
high-frequency limit, the driving field averages out, leading to an
effective uniform lattice that supports delocalized transport and coherent
Bloch-type oscillations governed by the underlying real or complex
dispersion.

At intermediate driving frequencies, a rich interplay emerges between
hopping asymmetry, harmonic confinement, and Floquet modulation. Nontrivial
quasi-energy band structures, hybridized localized--extended states, and
dynamically stabilized real spectra can appear even in the absence of $%
\mathcal{PT}$ symmetry. These regimes serve as the foundation for the
subsequent analysis of Hermitian and non-Hermitian Floquet dynamics
presented in Sec. \ref{Bloch oscillations}.

\section{Floquet Hamiltonian and Solutions}

\label{Floquet Hamiltonian and Solutions}

We now turn to the general case of a finite driving frequency $\omega $,
where the system exhibits genuine time-periodic dynamics. Equation (\ref%
{H_og}) defines a periodically driven lattice Hamiltonian, which can be
analyzed within the framework of Floquet theory. The essential idea is to
map a time-dependent periodic problem onto an equivalent eigenvalue problem
in an enlarged Hilbert space, known as the Sambe space, thereby transforming
the time evolution problem into a static one.

Considering a general time-periodic Hamiltonian
\begin{equation}
\mathcal{H}(t+T)=\mathcal{H}(t),
\end{equation}%
where $T=2\pi /\omega $ is the driving period. Mathematically, Floquet
theory is analogous to the Bloch theorem for spatially periodic crystals:
While spatial periodicity induces a band structure in momentum space,
temporal periodicity gives rise to the so-called Floquet quasi-energy bands
\cite{Shirley,Sambe}.

The solution to the time-dependent Schr\"{o}dinger equation can be expressed
as
\begin{equation}
|\psi _{l}(t)\rangle =e^{-i\epsilon _{l}t}|\phi _{l}(t)\rangle ,
\label{phi_b}
\end{equation}%
where $|\phi _{l}(t)\rangle $ is the $l$-th Floquet mode satisfying
periodicity $|\phi _{l}(t+T)\rangle =|\phi _{l}(t)\rangle $.

Throughout this section, we set $\hbar =1$ for simplicity. Substituting Eq. (%
\ref{phi_b}) into the Schr\"{o}dinger equation yields the Floquet eigenvalue
equation
\begin{equation}
\lbrack \mathcal{H}(t)-i\partial _{t}]|\phi _{l}(t)\rangle =\epsilon
_{l}|\phi _{l}(t)\rangle ,  \label{eq_fs}
\end{equation}%
which can be formally understood as a time-independent eigenvalue problem in
the extended Sambe space
\begin{equation}
\mathbf{S}=\mathbf{H}\otimes \mathbf{T},
\end{equation}%
where $\mathbf{H}$ is the Hilbert space of the original system, and $\mathbf{%
T}$ is the space of complex-valued periodic functions with period $T$.

Since both $\mathcal{H}(t)$ and $|\phi _{l}(t)\rangle $ are periodic, we can
expand them as
\begin{equation}
|\phi _{l}(t)\rangle =\sum_{m=-\infty }^{\infty }e^{-im\omega t}|\phi
_{l}^{(m)}\rangle ,  \label{F1}
\end{equation}%
and
\begin{equation}
\mathcal{H}(t)=\sum_{n=-\infty }^{\infty }e^{-in\omega t}\mathcal{H}^{(n)}.
\label{F2}
\end{equation}%
Substituting Eqs. (\ref{F1}) and (\ref{F2}) into Eq. (\ref{eq_fs}) yields
the following coupled equations:
\begin{equation}
\sum_{n}\left( \mathcal{H}^{(n-m)}-n\omega \delta _{n,m}\right) |\phi
_{l}^{(n)}\rangle =\epsilon _{l}|\phi _{l}^{(m)}\rangle ,
\end{equation}%
where $n$ and $m$ label the Fourier components. The corresponding Floquet
Hamiltonian matrix elements are given by
\begin{equation}
\mathcal{H}_{n,m}^{(F)}=\mathcal{H}^{(n-m)}-n\omega \delta _{n,m},
\end{equation}%
so that the complete Floquet Hamiltonian can be represented as
\begin{equation}
\mathcal{H}^{(F)}=%
\begin{pmatrix}
\ddots & \vdots & \vdots & \vdots &  \\
\cdots & \mathcal{H}^{(0)}+\omega & \mathcal{H}^{(-1)} & \mathcal{H}^{(-2)}
& \cdots \\
\cdots & \mathcal{H}^{(1)} & \mathcal{H}^{(0)} & \mathcal{H}^{(-1)} & \cdots
\\
\cdots & \mathcal{H}^{(2)} & \mathcal{H}^{(1)} & \mathcal{H}^{(0)}-\omega &
\cdots \\
& \vdots & \vdots & \vdots & \ddots%
\end{pmatrix}%
.
\end{equation}%
Each block corresponds to the coupling between different Fourier modes. The
diagonal terms $\mathcal{H}^{(0)}\pm n\omega $ represent the static part
shifted by integer multiples of the driving frequency, whereas the
off-diagonal blocks $\mathcal{H}^{(n-m)}$ describe the coupling between
different photon sectors.

In numerical simulations, the Floquet matrix must be truncated to a finite
number of photon sectors, i.e., $|m|\leq M$, leading to the approximate
block matrix
\begin{equation}
\mathcal{H}_{M}^{(F)}=%
\begin{pmatrix}
\mathcal{H}^{(0)}+M\omega & \mathcal{H}^{(-1)} & \cdots & \mathcal{H}^{(-2M)}
\\
\mathcal{H}^{(1)} & \mathcal{H}^{(0)}+(M-1)\omega & \cdots & \mathcal{H}%
^{(-2M+1)} \\
\vdots & \vdots & \ddots & \vdots \\
\mathcal{H}^{(2M)} & \mathcal{H}^{(2M-1)} & \cdots & \mathcal{H}%
^{(0)}-M\omega%
\end{pmatrix}%
.  \label{H_MF}
\end{equation}%
The Fourier components $\mathcal{H}^{(n)}$ for our quadratic potential model
are explicitly given by
\begin{equation}
\mathcal{H}^{(n)}=%
\begin{cases}
\sum_{l=-L}^{L}\frac{F_{0}l^{2}}{2}\,a_{l}^{\dagger }a_{l}, & n=\pm 1, \\%
[6pt]
-J\sum_{l=-L}^{L}(a_{l+1}^{\dagger }a_{l}+\mathrm{H.c.}), & n=0, \\[6pt]
0, & \text{otherwise}.%
\end{cases}%
\end{equation}%
Diagonalizing Eq. (\ref{H_MF}) yields the Floquet quasi-energies $\epsilon
_{l}$ and the corresponding Floquet modes $|\phi _{l}(t)\rangle $.

Alternatively, the Floquet Hamiltonian can be defined from the one-period
time-evolution operator,
\begin{equation}
U(T,0)=\mathcal{T}\exp \left[ -i\int_{0}^{T}\mathcal{H}(t)\,dt\right] ,
\end{equation}%
where $\mathcal{T}$ denotes time ordering. To evaluate $U(T,0)$ numerically,
the evolution period $T=2\pi /\omega $ can be divided into $Q$ small time
slices of width $\Delta t=T/Q$, such that
\begin{equation}
U_{q}=e^{-i\mathcal{H}(t)\Delta t},
\end{equation}%
and
\begin{equation}
U(T,0)=U_{Q}U_{Q-1}\cdots U_{2}U_{1},
\end{equation}%
where\textbf{\ }$q=1,2,...Q.$ The full-period evolution can be expressed in
terms of an effective static Floquet Hamiltonian $\mathcal{H}_{\mathrm{F}}$,
which satisfy
\begin{equation}
U(T,0)=\exp [-i\mathcal{H}_{\mathrm{F}}T],
\end{equation}%
leading to
\begin{equation}
\mathcal{H}_{\mathrm{F}}=\frac{i}{T}\ln [U(T,0)].  \label{FloquetH}
\end{equation}%
The eigenvalue problem of $\mathcal{H}_{\mathrm{F}}$,
\begin{equation}
\mathcal{H}_{\mathrm{F}}|\varphi _{l}\rangle =\epsilon _{l}|\varphi
_{l}\rangle ,  \label{HF}
\end{equation}%
provides the quasi-energies $\epsilon _{l}$ and Floquet eigenstates $%
|\varphi _{l}\rangle $. Generally, since $\mathcal{H}(t)$ at different times
do not commute, $U(T,0)$ cannot be evaluated analytically. However, for weak
driving amplitudes, one can apply time-dependent perturbation theory to
approximate $\mathcal{H}_{\mathrm{F}}$ and capture the leading-order
corrections to the quasi-energy spectra\ \cite{LCH}.

\section{Gauge Transformation and Analytical Theory}

\label{Gauge Transformation}

We now give a physically transparent and quantitatively controlled estimate
for the optimal frequency $\omega _{c}$ at which the Floquet spectrum
exhibits the best near-equally-spaced structure. To understand the origin of
$\omega _{c}$, it is convenient to remove the explicit on-site drive by a
time-dependent gauge transformation. We define
\begin{equation}
U(t)=\exp [-i\phi (t)\sum_{l=-L}^{L}l^{2}a_{l}^{\dagger }a_{l}],\phi (t)=%
\frac{F_{0}}{\omega }\sin (\omega t).  \label{eq:gauge_transform}
\end{equation}%
Then
\begin{equation}
\dot{\phi}(t)=F_{0}\cos (\omega t),
\end{equation}%
and the transformed Hamiltonian
\begin{equation}
\mathcal{H}_{g}(t)=U^{\dagger }(t)\mathcal{H}(t)U(t)-iU^{\dagger
}(t)\partial _{t}U(t)  \label{eq:Hg_def}
\end{equation}%
takes a particularly simple form. Indeed, the second term gives
\begin{eqnarray}
-iU^{\dagger }(t)\partial _{t}U(t) &=&-\dot{\phi}(t)\sum_{l=-L}^{L}l^{2}(|l%
\rangle \langle l|)  \notag \\
&=&-F_{0}\cos (\omega t)\sum_{l=-L}^{L}l^{2}(|l\rangle \langle l|),
\end{eqnarray}%
which exactly cancels the explicit time-dependent quadratic potential in Eq.
(\ref{H_og}).

Introducing the local basis
\begin{equation}
a_{l}^{\dagger }|0\rangle \equiv |l\rangle ,
\end{equation}%
where $|0\rangle $ is the vacuum state. Therefore the only remaining time
dependence is transferred into the hopping term
\begin{eqnarray}
\mathcal{H}_{g}(t) &=&-J\sum_{l=-L}^{L-1}[U^{\dagger }(t)(|l+1\rangle
\langle l|)U(t)  \notag \\
&&+U^{\dagger }(t)(|l\rangle \langle l+1|)U(t)].
\end{eqnarray}%
Using
\begin{equation}
U(t)|l\rangle =e^{-i\phi (t)l^{2}}|l\rangle ,
\end{equation}%
we obtain
\begin{equation}
U^{\dagger }(t)(|l+1\rangle \langle l|)U(t)=e^{-i\phi (t)\left[
(l+1)^{2}-l^{2}\right] }(|l+1\rangle \langle l|).
\end{equation}%
Since
\begin{equation}
(l+1)^{2}-l^{2}=2l+1,
\end{equation}%
the transformed Hamiltonian becomes
\begin{equation}
\mathcal{H}_{g}(t)=-J\sum_{l=-L}^{L-1}e^{-i\alpha _{l}\sin (\omega
t)}(|l+1\rangle \langle l|)+\mathrm{H.c.},  \label{eq:Hg_phase_dressed}
\end{equation}%
with the bond-dependent modulation index
\begin{equation}
\alpha _{l}=\frac{F_{0}}{\omega }(2l+1).  \label{eq:alpha_l_def}
\end{equation}

The original driven quadratic potential is exactly equivalent to a hopping
problem with bond-dependent periodic Peierls phases [seen in Eq. (\ref%
{eq:Hg_phase_dressed})]. The crucial feature is that the modulation strength
is not uniform in space; it grows with $|2l+1|$, and is therefore strongest
on the outermost bonds.

We now expand the phase factor by the Jacobi--Anger identity
\begin{equation}
e^{i\alpha _{l}\sin (\omega t)}=\sum_{m=-\infty }^{\infty }\mathcal{J}%
_{m}(\alpha _{l})e^{im\omega t},  \label{eq:Jacobi_Anger}
\end{equation}%
where $\mathcal{J}_{m}$ is the Bessel function of the first kind.
Substituting Eq. (\ref{eq:Jacobi_Anger}) into Eq. (\ref{eq:Hg_phase_dressed}%
), we obtain
\begin{equation}
\mathcal{H}_{g}(t)=-\sum_{m=-\infty }^{\infty }\sum_{l=-L}^{L-1}J\mathcal{J}%
_{m}(\alpha _{l})e^{-im\omega t}(|l+1\rangle \langle l|)+\mathrm{H.c.},
\label{eq:Hg_fourier}
\end{equation}%
this expression has a direct physical meaning: on bond $l$, the amplitude
for an $m$-photon-assisted hopping process is
\begin{equation}
J_{l}^{(m)}=J\mathcal{J}_{m}(\alpha _{l}).  \label{eq:Jl_m}
\end{equation}%
Hence the drive does not merely shift energies; it redistributes hopping
strength into different photon sectors, and this redistribution is strongly
bond dependent. We next explain why the first pronounced optimal ladder
structure should be controlled by the single-photon process on the outermost
bonds.

In the high-frequency regime, $\omega \gg F_{0}$, all $\alpha _{l}$ are
small, then we have
\begin{equation}
|\alpha _{l}|\text{ }=|\frac{F_{0}}{\omega }(2l+1)|\ll 1.
\end{equation}%
Then the Bessel functions satisfy
\begin{equation}
\mathcal{J}_{0}(\alpha _{l})\approx 1-\frac{\alpha _{l}^{2}}{4},\mathcal{J}%
_{1}(\alpha _{l})\approx \frac{\alpha _{l}}{2},\mathcal{J}_{m\geq 2}(\alpha
_{l})=O(\alpha _{l}^{m}).  \label{eq:Bessel_small_alpha}
\end{equation}%
Thus, as one lowers $\omega $ from very large values, the first nontrivial
photon-assisted process that becomes appreciable is the $m=1$ channel.
Higher-photon processes are parametrically smaller at this stage.

At the same time, because
\begin{equation}
|\alpha _{l}|\text{ }=\frac{F_{0}}{\omega }|2l+1|
\end{equation}%
increases monotonically with the distance from the chain center, the
outermost bonds reach the strong-mixing regime first. The maximal value of $%
|2l+1|$ is attained at
\begin{equation}
l=L-1\quad \text{or}\quad l=-L,
\end{equation}%
for which
\begin{equation}
|2l+1|_{\max }=2L-1.
\end{equation}%
Therefore the largest modulation index in the whole chain is
\begin{equation}
\alpha _{\mathrm{edge}}=\frac{(2L-1)F_{0}}{\omega }.  \label{eq:alpha_edge}
\end{equation}%
This leads to a natural physical picture: as $\omega $ is lowered from the
deep high-frequency regime, the first strong rearrangement of Floquet levels
is triggered when the one-photon sideband weight on the outermost bonds
becomes maximal. Since the best near-equally-spaced quasienergy structure is
precisely the point where different Floquet sectors are mixed most
efficiently and reorganize into an approximate ladder after zone folding,
the first optimal frequency should be estimated by maximizing $\left\vert
\mathcal{J}_{1}(\alpha _{\mathrm{edge}})\right\vert $.

Let $\alpha _{c}$ denote the first positive extremum of $|\mathcal{J}%
_{1}(\alpha )|$. It is determined by
\begin{equation}
\frac{d}{d\alpha }\mathcal{J}_{1}(\alpha )=0.
\end{equation}%
Using the standard identity
\begin{equation}
\frac{d}{d\alpha }\mathcal{J}_{1}(\alpha )=\frac{\mathcal{J}_{0}(\alpha )-%
\mathcal{J}_{2}(\alpha )}{2},
\end{equation}%
the extremum condition becomes
\begin{equation}
\mathcal{J}_{0}(\alpha _{c})=\mathcal{J}_{2}(\alpha _{c}).
\end{equation}%
Its first positive solution is
\begin{equation}
\alpha _{c}\approx 1.841.  \label{eq:x1star}
\end{equation}%
Imposing the condition
\begin{equation}
\alpha _{\mathrm{edge}}=\alpha _{c},
\end{equation}%
namely
\begin{equation}
\frac{(2L-1)F_{0}}{\omega _{c}}=\alpha _{c},
\end{equation}%
we finally obtain the analytical estimate
\begin{equation}
\omega _{c}\approx \frac{(2L-1)F_{0}}{\alpha _{c}}.  \label{eq:omega_c_main}
\end{equation}%
We emphasize again that (\ref{eq:omega_c_main}) does not describe a simple
resonance between two bare eigenvalues of the static Hamiltonian $H_{0}$.
Rather, it characterizes the point where the photon-assisted hopping induced
by the drive is strongest on the most strongly modulated bonds. At this
point, the coupling between neighboring Floquet sectors is maximized most
efficiently, so that after folding quasienergies back into the first Floquet
Brillouin zone, the spectrum is reorganized into the best approximate ladder
structure.

\section{Dynamics: Bloch-like oscillations and frequency-tuned Floquet
spectra}

\label{Bloch oscillations}

Bloch oscillation represents one of the most fascinating manifestations of
coherent quantum dynamics in periodic potentials \cite{Zener}. It has been
extensively explored in various ultracold atomic systems, including
degenerate Bose/Fermi gases \cite{Raithel,CG}, strongly correlated atomic
lattices \cite{Sherson,Pedersen}, and Bose--Einstein condensates \cite%
{Morsch,Chin,Cronin,Krzy}. Motivated by these seminal results, we analyze
the dynamical behavior of the time-periodically driven quadratic Hamiltonian
introduced in Sec. \ref{Model Hamiltonian}, and demonstrate the optimal
frequency $\omega _{c}$ for the Floquet spectrum with the best
near-equally-spaced structure by analytical methods in Sec. \ref{Gauge
Transformation}.

As shown above, the time-periodic Hamiltonian can be mapped to an effective
static Floquet problem. The structure of the Floquet quasi-energy spectrum
depends sensitively on the drive frequency $\omega $. In certain parameter
regimes and at specific drive frequencies, the Floquet Hamiltonian $\mathcal{%
H}_{\mathrm{F}}$ is found to support nearly equidistant quasi-energy
ladders. Such ladders directly imply periodic dynamics with well-defined
revival times and are responsible for Bloch-oscillation--like behavior in
the driven system.

\subsection{Frequency dependence of the Floquet spectrum}

\label{Sec:FrequencyDependence}

The Floquet problem can be formulated as an eigenvalue equation in the Sambe
space,
\begin{equation}
\mathcal{H}^{(F)}_{n,m}=\mathcal{H}^{(n-m)}-n\omega\delta_{n,m},
\end{equation}
where $\mathcal{H}^{(n-m)}$ are the Fourier components of $\mathcal{H}(t)$
and $n,m$ label the photon sectors.

In the high-frequency regime ($\omega \gg J$, $F_{0}L^{2}$), the
off-diagonal couplings between different photon sectors are strongly
suppressed, and the leading-order Floquet Hamiltonian reduces to its
time-averaged form $\mathcal{H}_{\mathrm{eff}}\approx H_{0}$. The
corresponding quasi-energy spectrum therefore reproduces that of the static
hopping Hamiltonian---an extended Bloch-band dispersion $E(k)=-2J\cos k$ for
the Hermitian case, or a complex dispersion $E(k)=-J_{1}e^{ik}-J_{2}e^{-ik}$
for non-Hermitian case.

As $\omega $ decreases to values comparable to the characteristic level
spacing of the confined static system, hybridization between neighboring
photon sectors mediated by $H^{(\pm 1)}$ becomes significant. In this
intermediate-frequency regime, resonant mixing reorganizes portions of the
Floquet spectra into nearly equidistant quasi-energy ladders. This occurs
when the spacing between a family of eigenstates of the static Hamiltonian
approaches an integer multiple of $\omega $, causing replicas from adjacent
photon sectors to align and hybridize. The resulting hybridized manifold can
be effectively described by an emergent Hamiltonian whose spectrum forms a
harmonic-like ladder within a finite quasi-energy window. The ladder spacing
$\Delta E$ is set by the residual hybridization splittings and remains
nearly constant across that subspace.

These emergent ladders typically arise in finite systems and manifest as
frequency windows where the distribution of nearest-neighbor level spacings
\begin{equation}
s_{l}\equiv \epsilon _{l+1}-\epsilon _{l},
\end{equation}%
become narrowly peaked around a constant $\Delta E$. A convenient
quantitative diagnostic is the normalized variance,
\begin{equation}
\Delta (\omega )=\frac{\sum_{l}(s_{l}-\overline{s})^{2}}{2L\overline{s}^{2}},
\label{eq:variance}
\end{equation}%
evaluated over a contiguous subset of levels ($l=-L,\dots ,L-1$), where $%
\overline{s}=\sum_{l}\frac{1}{2L}s_{l}$. Small values $\Delta (\omega )\ll 1$
indicate the emergence of nearly equidistant quasi-energy ladders.

\begin{figure*}[tbp]
\centering
\includegraphics[bb=57 64 2377 1123, width=18 cm, clip]{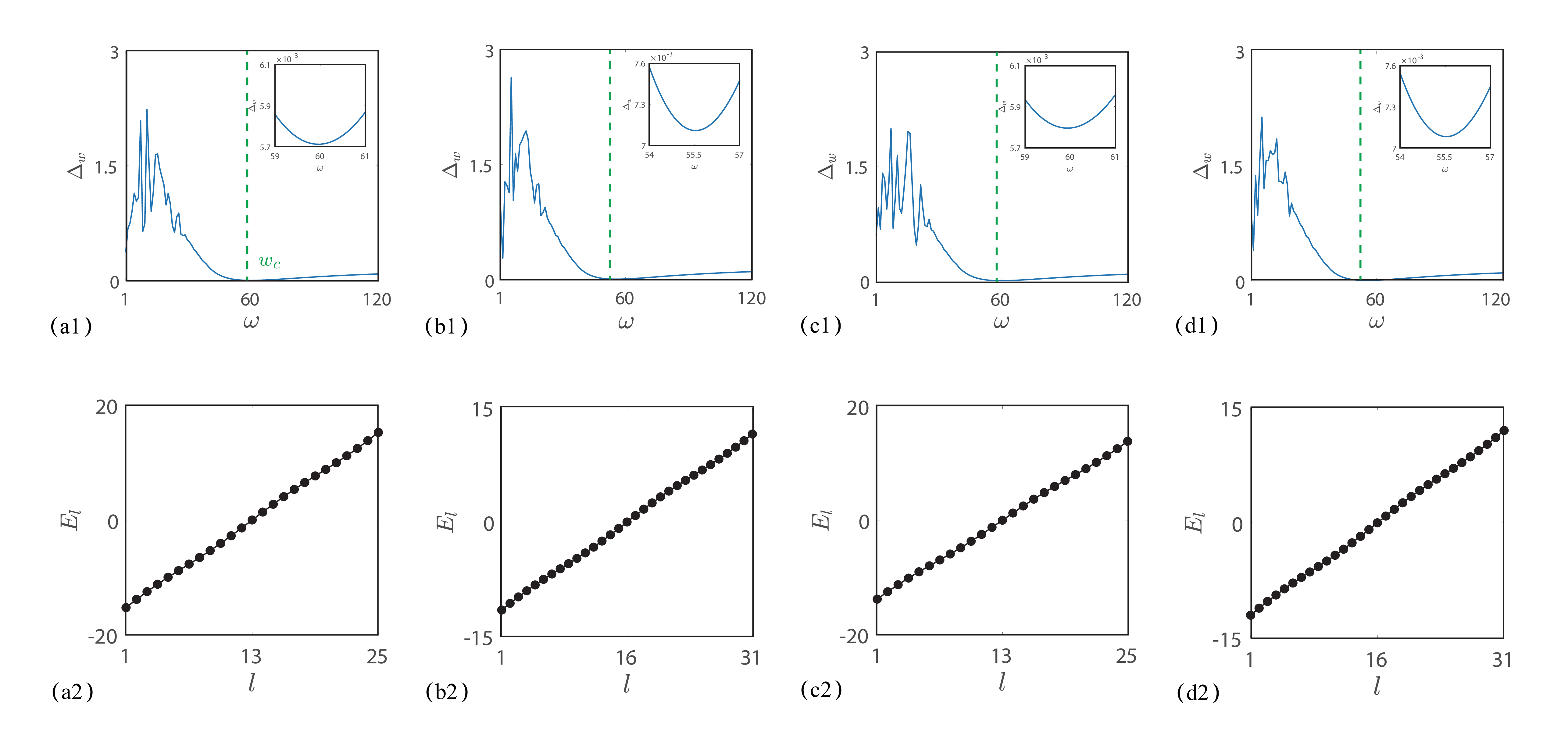}
\caption{{}Panels (a1--d1) display the numerical results for the normalized
variance $\Delta (\protect\omega )$ of the quadratic Floquet systems as a
function of the Floquet driving frequency $\protect\omega $ with two
Hermitian cases ($J_{1}=J_{2}$) and two non-Hermitian cases ($J_{1}\neq
J_{2} $). Panels (a2--d2) display the corresponding Floquet quasi-energy
spectra at the critical frequency $\protect\omega _{c}$ (indicated by green
dashed lines) where $\Delta (\protect\omega )$ attains its minimum $\Delta (%
\protect\omega _{c})$. We note that the quasi-energy spectrum becomes
approximately equidistant, forming a Floquet ladder structure when $\protect%
\omega \approx \protect\omega _{c}$. The numerical results $\protect\omega %
_{c}$ are in good agreement with the analytical results obtained from the
gauge transformation in Sec. \protect\ref{Gauge Transformation}.
Specifically: panels (a1--a2) $J_{1}=J_{2}=8$, $F_{0}=4$, $L=12$; panels
(b1--b2) $J_{1}=J_{2}=6$, $F_{0}=3$, $L=15$; panels (c1--c2) $J_{1}=7$, $%
J_{2}=7.5$, $F_{0}=4$, $L=12$; panels (d1--d2) $J_{1}=6$, $J_{2}=6.5$, $%
F_{0}=3$, $L=15$. The step size of frequency $\protect\omega $\ is $0.05$.}
\label{fig3}
\end{figure*}

Numerically, we truncate the photon index to $|n|\leq M$ and diagonalize the
finite block matrix $\mathcal{H}_{M}^{(F)}$ to obtain the quasi-energies $%
\{\epsilon _{l}(\omega )\}$. Scanning $\omega $ reveals the generic
behaviors described above. Figure \ref{fig3} show the normalized variance $%
\Delta (\omega )$ and the corresponding Floquet spectra for both Hermitian
cases ($J_{1}=J_{2}$) and non-Hermitian cases ($J_{1}\neq J_{2}$) quadratic
Floquet systems. Panels (a1--d1) display $\Delta (\omega )$ as a function of
$\omega $, while Panels (a2--d2) show the corresponding Floquet quasi-energy
spectra at the critical frequency $\omega _{c}$ (indicated by green dashed
lines), where $\Delta (\omega )$ attains its minimum $\Delta (\omega _{c})$.
At these frequencies, the quasi-energy spectra become approximately
equidistant, forming Floquet ladders structure. The numerical results $%
\omega _{c}$ are in good agreement with the analytical results obtained from
the gauge transformation in Sec. \ref{Gauge Transformation}.

When $\omega $ is tuned across values where uncoupled photon-sector energies
would intersect, off-diagonal couplings open avoided crossings that can
either spoil or enhance equidistance depending on matrix-element symmetries.
Finite system size and photon truncation $M$ also play important roles:
Ladder formation is most robust when the relevant states are well-contained
within the truncated Floquet Hilbert space and higher-order photon processes
remain negligible or act uniformly across them.

Building upon these results, we now analyze the two limiting regimes of the
Floquet frequency: In the low-frequency limit ($\omega \rightarrow 0$), the
Hamiltonian in Eq. (\ref{H_og}) effectively reduces to $H_{1}=H_{0}+H(0)$.
The eigenstates of this static system consist of two characteristic
families, a low-energy harmonic-oscillator-like set and a high-energy
Wannier-Stark-localized set \cite{Ali}. The overall state localization is
therefore strong, dominated by the confining quadratic potential. In the
high-frequency limit ($\omega \rightarrow \infty $), the potential
oscillates rapidly, such that its time-averaged contribution vanishes and
the effective Hamiltonian reduces to $H_{0}$. The eigenstates are extended
Bloch waves, corresponding to a delocalized transport regime.

When a contiguous block of Floquet quasi-energies form an approximately
uniform ladder,
\begin{equation}
E_{l}(\omega )\simeq E_{0}(\omega )+l\Delta E(\omega ),
\end{equation}%
an arbitrary initial state decomposed in these eigenstates,
\begin{equation}
|\varphi (0)\rangle =\sum_{l}c_{l}|\varphi _{l}\rangle ,
\end{equation}%
evolves as
\begin{equation}
|\varphi (t)\rangle =e^{-iE_{0}t}\sum_{l}c_{l}e^{-il\Delta Et}|\varphi
_{l}\rangle .
\end{equation}%
Consequently the system exhibits revivals at times
\begin{equation}
t_{c}=\frac{2\pi c}{\Delta E(\omega )},\text{ }c\in \mathbb{Z},
\end{equation}%
and the fidelity $\mathcal{F}(t)=|\langle \varphi (0)|\varphi (t)\rangle |$
displays sharp periodic peaks. Importantly, $\Delta E(\omega )$ is a
function of the drive frequency: By scanning $\omega $ and extracting $%
\Delta E(\omega )$ from the quasi-energy spectrum (or directly from Fourier
analysis of the time dynamics), one can identify optimal drive frequencies
that maximize ladder uniformity and fidelity revival contrast.

\subsection{Non-Hermitian Floquet ladders and emergent Hermiticity}

For non-Hermitian $H_{0}^{\prime }$ (e.g., $J_{1}\neq J_{2}$), the Floquet
quasi-energies are generally complex. Nevertheless, the effective Floquet
Hamiltonian $\mathcal{H}_{\mathrm{F}}^{\prime }$ can develop partially real
quasi-energy ladders under suitable driving conditions. This dynamical
stabilization arises when non-Hermitian components either decouple from the
photon sectors forming the ladder, or when symmetric hybridization among
these sectors effectively cancels the net gain and loss within that
subspace. Such behavior depends sensitively on system parameters and becomes
most transparent in truncated Floquet spectra.

When a subset of quasi-energies $\{E_{l}^{\prime }\}$ are nearly real and
equidistant,
\begin{equation}
E_{l}^{\prime }(\omega )\simeq E_{0}^{\prime }(\omega )+l\Delta E^{\prime
}(\omega ),
\end{equation}%
initial states with dominant support on this subspace exhibit Hermitian-like
revivals with a characteristic period $t_{c}^{\prime }=2\pi /\Delta
E^{\prime }(\omega )$. The fidelity, computed via the biorthogonal inner
product (or the usual inner product when the subspace is effectively
Hermitian), displays periodic peaks whose visibility is determined by the
small imaginary parts of the quasi-energies.

By tuning the driving frequency $\omega $, one can directly reshape the
Floquet quasi-energy landscape. At large $\omega $, the drive averages out
and the dynamics follow the static dispersion; near resonant frequency,
hybridization between Floquet replicas produces emergent, nearly equidistant
ladders accompanied by Bloch-type revivals. In non-Hermitian systems,
suitable parameter choices can further stabilize partially real ladders that
sustain Hermitian-like oscillations. The combined use of spectral-spacing
variance and real-time propagation provides a consistent framework for
identifying and exploiting frequency windows that host nearly perfect
revivals.

\begin{figure*}[tbp]
\centering
\includegraphics[bb=70 45 2173 941, width=18 cm, clip]{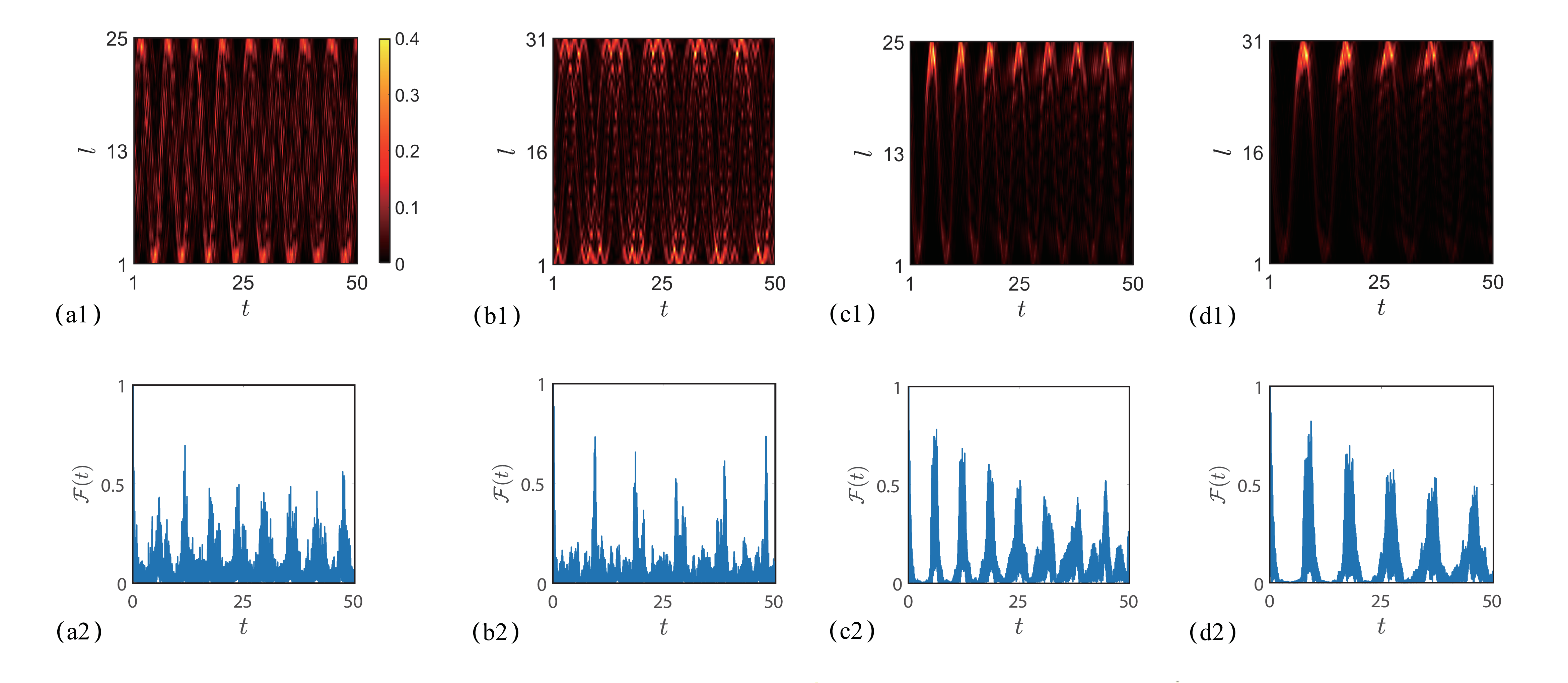}
\caption{{}Panels (a1--d1) display the numerical results for the evolution
states $|\protect\varphi (t)\rangle $ of the quadratic Floquet systems as a
function of the evolution time $t$ with two Hermitian cases ($J_{1}=J_{2}$)
and two non-Hermitian cases ($J_{1}\neq J_{2}$). Panels (a2--d2) display the
corresponding fidelity $\mathcal{F}(t)$ between the evolved state and
initial state. We note that the numerical results reveal clear Bloch-like
oscillatory revivals whose periods agree well with the analytical
expressions $t_{c}=2\protect\pi /\Delta E(\protect\omega )$ [panels
(a1--b1)] and $t_{c}^{\prime }=2\protect\pi /\Delta E^{\prime }(\protect%
\omega )$ [panels (c1--d1)]. The numerical results of the fidelity exhibit
the same periodicity as the revival dynamics, confirming the correspondence
between quasi-energy ladder spacing and dynamical recurrence. The minimum
periods satisfy $t_{1}\simeq 5.76$, $9.45$, $6.28$, and $9.15$ in panels
(a1-d1), respectively, all of which are in excellent agreement with the
analytical predictions. The initial state is prepared as $|\protect\varphi %
(0)\rangle =\sum_{l=L-4}^{L+6}11^{-1/2}|\protect\varphi _{l}\rangle $ with
time step $\Delta t=0.01$, respectively. Other parameters are identical to
those in Fig. \protect\ref{fig3}. The colorbar indicates the probabilities
at each position in the time-evolved states.}
\label{fig4}
\end{figure*}

Figures \ref{fig4}(a1--d1) illustrates the time evolution of states in
Hermitian cases ($J_{1}=J_{2}$) and non-Hermitian cases ($J_{1}\neq J_{2}$)
quadratic Floquet systems under periodic boundary conditions at the critical
frequency $\omega _{c}$. The numerical results reveal clear Bloch-like
oscillatory revivals whose periods agree well with the analytical
expressions $t_{c}=2\pi /\Delta E(\omega )$ [Panels (a1--b1)] and $%
t_{c}^{\prime }=2\pi /\Delta E^{\prime }(\omega )$ [Panels (c1--d1)]. The
initial state is prepared as $|\varphi (0)\rangle
=\sum_{l=L-4}^{L+6}11^{-1/2}|\varphi _{l}\rangle $, with time step $\Delta
t=0.01$, respectively. Panels (a2--d2) display the fidelity $\mathcal{F}(t)$
between the evolved state and the initial state, which exhibits the same
periodicity as the revival dynamics, confirming the correspondence between
quasi-energy ladder spacing and dynamical recurrence. All of these results
are in excellent agreement with the analytical predictions.

To identify frequency windows where ladder formation is robust and to verify
the resulting revival dynamics in more realistic settings, we also
investigate the effect of quenched disorder in the hopping amplitudes.
Concretely, we replace the uniform right(left) hoppings $J_{1}$($J_{2}$)\ by
site-dependent values
\begin{equation}
J_{1}^{(l)}=J_{1}+\mathrm{ran}(-\lambda ,\lambda )\text{, }J_{2}^{(l)}=J_{2}+%
\mathrm{ran}(-\lambda ,\lambda ),
\end{equation}%
where $\mathrm{ran}(-\lambda ,\lambda )$ denotes a uniform random number in
the interval $(-\lambda ,\lambda )$. The driven disordered Hamiltonian then
reads
\begin{eqnarray}
\mathcal{H}(t) &=&-\sum_{l=-L}^{L}[J_{1}^{(l)}a_{l+1}^{\dagger
}a_{l}+J_{2}^{(l)}a_{l}^{\dagger }a_{l+1}]  \notag \\
&&+F_{0}\sum_{l=-L}^{L}l^{2}\cos (\omega t)a_{l}^{\dagger }a_{l}.
\end{eqnarray}

\begin{figure}[tbp]
\centering
\includegraphics[bb=73 41 1164 498, width=8.0cm, clip]{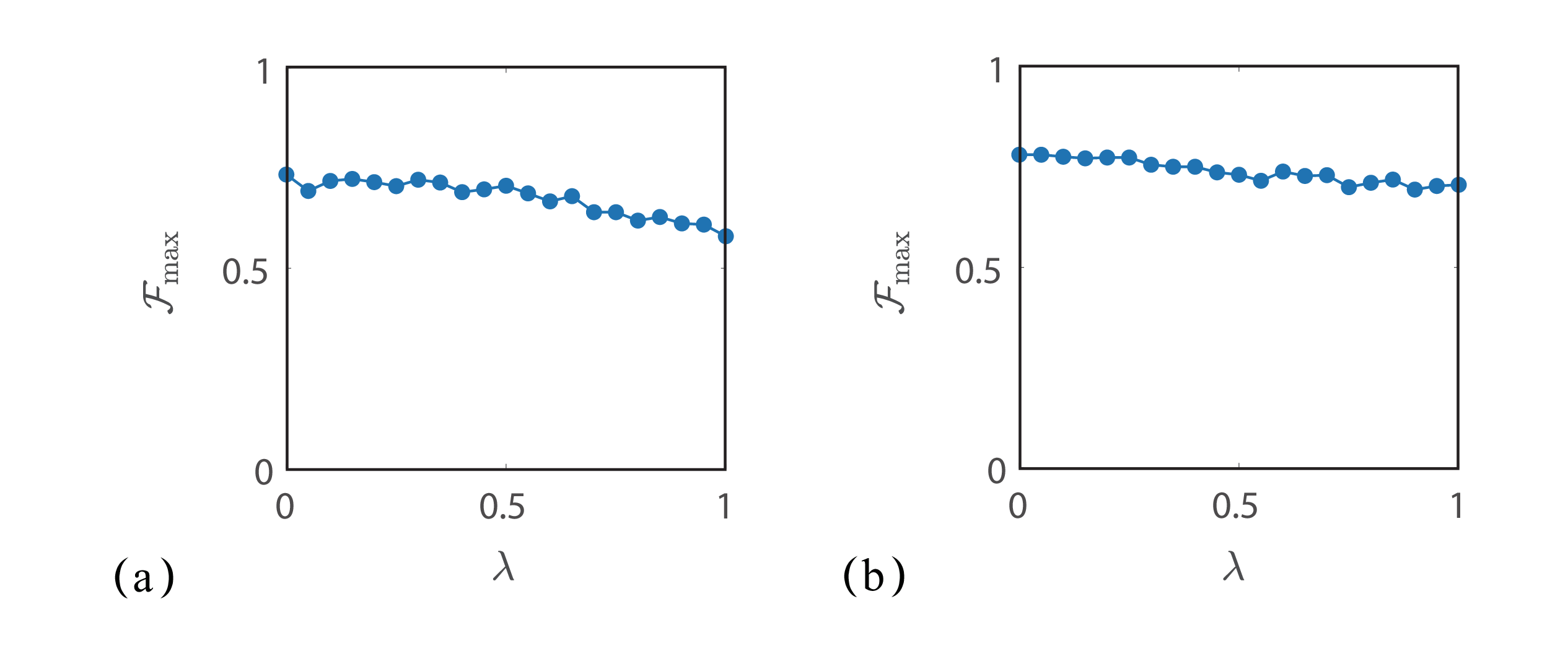}
\caption{Maximum fidelity peak $\mathcal{F}_{\max }$ as a function of the
disorder strength $\protect\lambda $ for panel (a) Hermitian ($J_{1}=J_{2}=6$%
, $F_{0}=3$, $L=15$) and panel (b) non-Hermitian ($J_{1}=7$, $J_{2}=7.5$, $%
F_{0}=4$, $L=12$) parameter sets. Each point represents an average of the
maximum fidelity peak over an ensemble of disorder realizations (see main
text). The results show that Bloch-like revivals at the near-resonant
frequency $\protect\omega _{c}$ persist up to moderate disorder. All other
parameters follow Figs. \protect\ref{fig3}.}
\label{fig5}
\end{figure}

Figure \ref{fig5} summarizes the key numerical findings. For each value of
the disorder strength $\lambda $, we compute the time evolution at the
near-resonant frequency $\omega _{c}$ identified and record the maximum
fidelity peak $\mathcal{F}_{\max }$ associated with periodic revivals (this
peak measures the largest return overlap occurring at the predicted revival
time $t_{c}=2\pi /\Delta E$ and its multiples). To suppress sample-to-sample
fluctuations, we average $\mathcal{F}_{\max }$ over many disorder
realizations (in practice, we use a moderate ensemble, e.g. twenties of
realizations, to obtain smooth curves).

The numerical results demonstrate that the Floquet-induced revival dynamics
are robust against moderate amounts of static disorder. As $\lambda $
increases from zero, $\mathcal{F}_{\max }$ remains large for a finite
disorder window, indicating that the quasi-equidistant ladder and the
coherent revivals survive spatial inhomogeneities. Beyond a characteristic
disorder scale, revivals gradually deteriorate: The peak fidelity decreases
and revival contrast is lost, reflecting (i) disorder-induced broadening and
fragmentation of the ladder subspace, and (ii) enhanced dephasing among the
ladder eigenvalues. Physically, this robustness can be understood because
ladder formation relies on local resonant hybridization in energy (Floquet
replica alignment) and on hybridization-induced emergent subspaces; moderate
spatial randomness only weakly perturbs these energy space resonances, while
strong disorder eventually destroys the coherent structure.

In the non-Hermitian case, the qualitative picture remains similar, although
the quantitative stability depends sensitively on the choice of parameters
(asymmetry hopping strength, driving strength, system size, and truncation
of the Floquet sectors). In some parameter regimes the periodic drive
continues to stabilize a near-real, uniformly spaced quasi-energy ladder up
to disorder strengths comparable to the Hermitian case; in others, the
interplay of nonreciprocity and disorder speeds the loss of revival
fidelity. This parameter dependence highlights the need for combined
spectral diagnostics (spacing variance ) together with direct dynamical
probes (fidelity) to assess robustness in any concrete experimental
realization.

Overall, these results indicate that Floquet-engineered quasi-energy ladders
and their associated Bloch-like revivals are not purely fine-tuned artifacts
of perfectly clean models: They persist under experimentally relevant levels
of static disorder, and their degradation with increasing $\lambda $ follows
clear spectral and dynamical signatures that can be monitored and, in some
cases, mitigated by adjusting drive parameters.

\section{Summary and Outlook}

\label{Summary}

We have investigated the quantum dynamics of a one-dimensional tight-binding
lattice subjected to a spatially quadratic and time-periodic potential,
focusing on both Hermitian and non-Hermitian hopping configurations. Within
the framework of Floquet theory, we map the explicitly time-dependent
Hamiltonian to an effective static Floquet operator and perform a systematic
analysis of its quasi-energy spectra and Floquet eigenstates as functions of
the driving frequency. Our central finding is the emergence of nearly
equidistant quasi-energy ladders at a set of critical driving frequencies,
signaled by a pronounced minimum in the normalized variance of the
nearest-neighbor level spacings. Analytical expressions for the critical
frequencies are derived via a gauge transformation.

The appearance of these quasi-harmonic ladders leads directly to long-time
periodic dynamics and high-contrast revival behavior of initially localized
wave packets. Numerical simulations confirm that such Bloch-like oscillatory
revivals occur in both Hermitian and non-Hermitian regimes, demonstrating
that a suitable periodic drive can dynamically stabilize an almost uniformly
spaced and nearly real quasi-energy spectrum even in the presence of
asymmetric hopping. This reveals a mechanism of Floquet-assisted dynamical
stabilization that persists outside the conventional Hermitian paradigm.

To assess the robustness of this mechanism, we introduce spatial disorder
into the hopping amplitudes and analyze the resulting Floquet dynamics. We
find that the near-resonant revival behavior remains stable over a finite
range of disorder strengths in both the Hermitian and non-Hermitian cases.
The maximum fidelity peak associated with multiple revival cycles decreases
only gradually as disorder increases, demonstrating that the emergent
quasi-energy ladder is not a fine-tuned feature of the clean system. Only
beyond a characteristic disorder scale does the ladder structure become
progressively distorted, leading to dephasing and the eventual suppression
of coherent revivals. This robustness underscores that Floquet engineering
of quasi-harmonic ladders provides a practical and disorder-tolerant route
for controlling coherent dynamics in driven lattice systems.

Overall, our work identifies a unifying mechanism for the formation of
equidistant quasi-energy structures in both Hermitian and non-Hermitian
tight-binding lattices under quadratic periodic driving, establishes clear
spectroscopic and dynamical signatures of this phenomenon, and demonstrates
its stability against realistic levels of disorder. These results broaden
the scope of Floquet engineering and point to new possibilities for
realizing controllable, long-lived dynamical states in driven quantum
platforms.

\acknowledgments We acknowledge the support of the Science \& Technology
Development Fund of Tianjin Education Commission for Higher Education(No.
2024KJ060).

\end{document}